\documentclass[aps,prb,jmp,amsmath,amssymb,reprint]{revtex4-1}
\usepackage{graphicx}% Include figure files
\usepackage{dcolumn}% Align table columns on decimal point
\usepackage{bm}% bold math
\usepackage{graphicx}
\usepackage{subfigure}
\usepackage{physics}
\begin{document}

%\preprint{AIP/123-QED}

\title{Superior thermoelectric performance via ``anti-reflection'' enabled double-barrier structures }% Force line breaks with \\
%\thanks{Footnote to title of article.}

\author{Swarnadip Mukherjee}
% \altaffiliation[Also at ]{Physics Department, XYZ University.}%Lines break automatically or can be forced with \\
\author{Pankaj Priyadarshi}
% \altaffiliation[Also at ]{Physics Department, XYZ University.}%Lines break automatically or can be forced with \\
\author{Bhaskaran Muralidharan}%
\email{bm@ee.iitb.ac.in}
\affiliation{Department of Electrical Engineering, Indian Institute of Technology Bombay, Powai, Mumbai-400076, India
}

%\author{C. Author}
% \homepage{http://www.Second.institution.edu/~Charlie.Author.}
%\affiliation{%
%Second institution and/or address%\\This line break forced% with \\
%}%

\date{\today}% It is always \today, today,
%  but any date may be explicitly specified

\begin{abstract}
	We demonstrate theoretically using the atomistic non-equilibrium Green's function formalism with the inclusion of self-consistent charging, the design of a superior thermoelectric generator based on an ``anti-reflection'' coated double barrier resonant tunnelling diode. Unlike a typical double barrier device, we show that enabling the anti-reflection design facilitates a ``boxcar'' type feature in its transmission spectrum, which significantly enhances the thermoelectric performance. It is demonstrated that the best operating regime of this device offers a maximum power in the range of $0.7$ to $0.9 MW/m^{2}$ at efficiencies ranging from $46$ to $54\%$ of Carnot efficiency. The physics of charge and heat transport in the ballistic regime of operation helps us gain additional insights on how a large number of transverse current carrying modes boost the output power and simultaneously how the diminishing effects of high-energy parasitic currents aid the efficiency. Finally, a comparative study with a conventional double barrier thermoelectric is presented in terms of standard performance parameters which clearly reveals the performance benefits of enabling an anti-reflection coating.
\end{abstract}

%\pacs{Valid PACS appear here}% PACS, the Physics and Astronomy
% Classification Scheme.
%\keywords{Suggested keywords}%Use showkeys class option if keyword
%display desired
\maketitle
Nano-structured solid state devices \cite{Dressel1,Dressel2,Dressel3,Mahan,Harman,Snider08,Poudel} with lineshape engineered densities of states (DOS) \cite{Dressel1,Dressel2,Dressel3,Mahan,Heremans08,Naka} have become promising candidates  for thermoelectric heat engines and refrigerators. An important direction in this context is the engineering of zero-dimensional confinement \cite{Kim09,Sothmann_Review} to obtain a delta shaped transmission profile \cite{Mahan}, which results in a maximum electronic figure of merit $zT$. However, for such a perfect energy filter, thermoelectric operation under open circuit conditions reaches the Carnot's limit \cite{Humphrey} yielding zero power output, thereby pointing to the inadequacy of using only $zT$ as a metric to analyze thermoelectric performance.\\
\indent In recent times, thermoelectric analysis based on output power and efficiency \cite{Naka,Esposito1,Esposito2,Esposito3,whitney,BMgriffoni,Sothmann_Review,Sothmann,akshaybm} has gained precedence, since $zT$ represents only the maximum efficiency point and its use as a sole metric does not provide a detailed picture of the actual device operation \cite{whitney,Whitney2015,Hershfield}. It was established recently \cite{whitney} that a finite width transmission window with a sharp transition profile in the shape of a "boxcar" proves to be more proficient in enhancing the efficiency at a given output power \cite{whitney,Whitney2015,Hershfield}. Based on the proposal of a thermoelectric generator \cite{Sothmann} using a hot cavity coupled to two cold junctions through a double barrier resonant tunnelling diode (RTD) structure, a realistic RTD thermoelectric\cite{akshaybm} was analyzed recently. It is clearly seen that such devices \cite{akshaybm} cannot generate significant amount of power at a high efficiency due to the sharp transmission profile. In another recent work \cite{Karbaschi2016}, a superlattice thermoelectric with an anti-reflection coating (ARC) demonstrated a better power-efficiency trade-off, via a well engineered "boxcar" transmission spectrum. However, the proposal suffers from multiple low transmission peaks that reduce the transmissivity considerably, and demands the use of a greater number of superlattice periods. \\
%\begin{center}
\begin{figure}[!htb]
	
	\subfigure[]{\includegraphics[height=0.15\textwidth,width=0.25\textwidth]{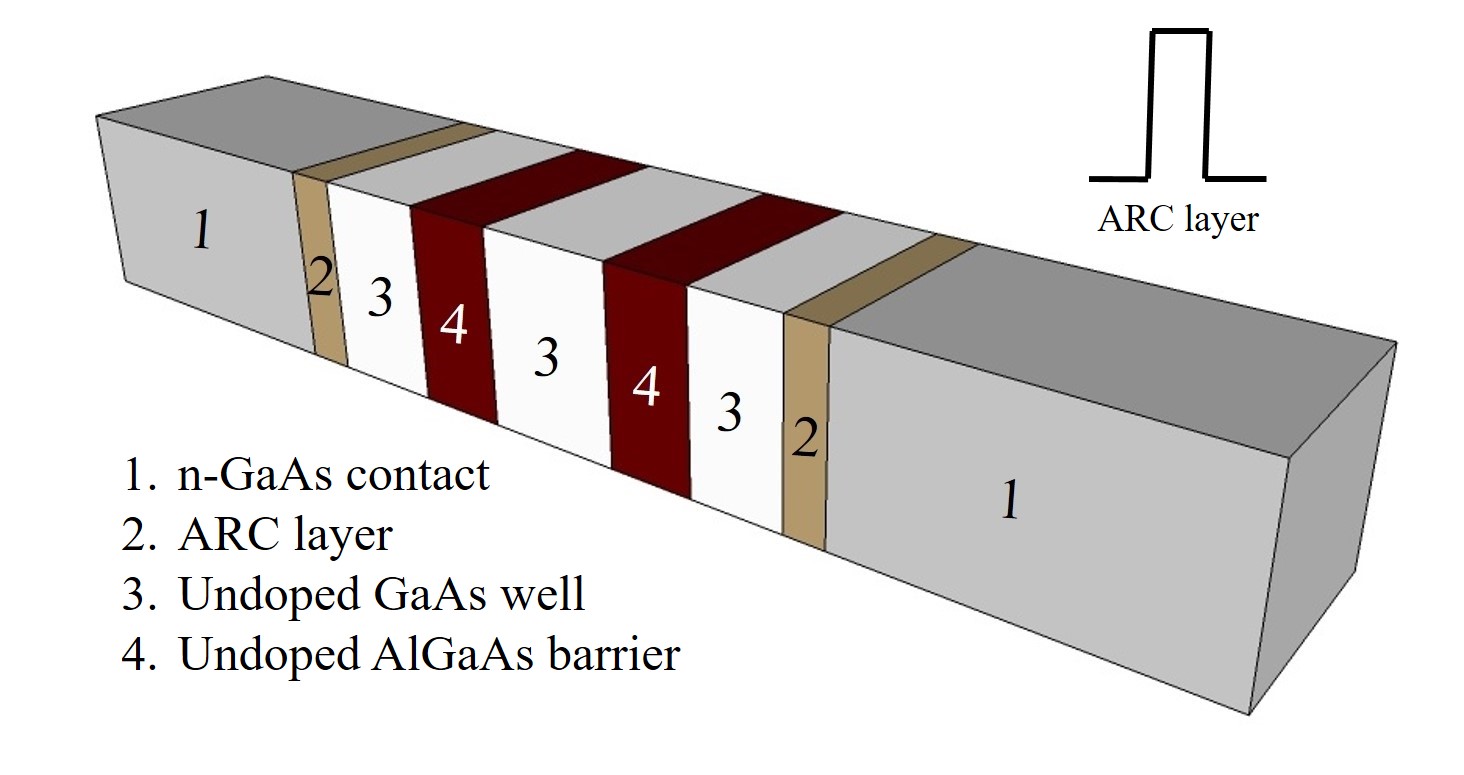}\label{1a}}
	\quad
	\subfigure[]{\includegraphics[height=0.15\textwidth,width=0.25\textwidth]{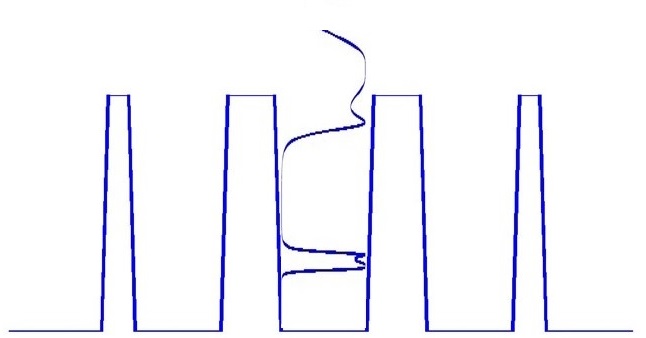}\label{1b}}
	\quad
	\caption{Device schematics: (a) Schematic of the ARC-RTD thermoelectric device that consists of two n-doped GaAs layers that form the contacts, while the undoped central RTD region, formed by a GaAs quantum well (white) and the AlGaAs barriers (dark red), is sandwiched between two ARC layers (brown). (b) Conduction band diagram of the device is shown along with the overall transmission spectrum. Enabling the ARC results in a unity transmission window of a finite width (almost similar to a boxcar function) with a very small ripple, unlike a regular RTD structure which produces a resonant sharp transmission peak.}
	\label{1}
\end{figure}
%\end{center}
\indent In this letter, using the atomistic non-equilibrium Green's function (NEGF) formalism with the inclusion of self-consistent charging, we propose a new bulk thermoelectric device design that consists of a conventional double-barrier RTD structure in between two ARC layers. Motivated by the fact that a finite width boxcar transmission spectrum enhances the efficiency at a significant finite power output, the ARC layer is designed such that the effective transmission in the neighborhood of main RTD resonance peak is increased, thereby improving the transmissivity in the energy range of interest. It was proposed earlier that the addition of ARC layers of same well width and half the barrier width as that of the central periodic heterostructure on both sides, improves the matching between the input and the central channel \cite{Iogansen}. Here, ARC layers act as an impedance transformer that matches the input to the load, analogous to microwave impedance matching. Design and theoretical investigations have also been carried out on the application of ARC on both sides of superlattice to form an electronic band-pass filter\cite{Martorell2004,Morozov2002} which improves the miniband transport\cite{Pacher2001}. \\
\indent Ideally, the ARC layer is designed to improve the transmission at a particular energy (say, $E_{m}$), however, a significant improvement is also observed in the neighborhood of that energy. The advantage of the ARC layers can be best understood from the physics behind it, which is based on two key points: (a) It should be a Bragg Reflector at that energy and (b) The potential profile of the ARC layer should be such that the electronic state at that energy becomes a Bloch eigenstate of the central region. These two conditions determine the transmission matrix elements of the ARC section in terms of the central region design parameters. However, there is no unique design solution as the different combinations of design parameters can lead to identical results. In our work, we incorporate the design criteria of ARC layers as proposed in Ref. 23. \\
\indent We will thus present a detailed quantitative study of the performance of an ARC enabled RTD thermoelectric (ARC-RTD) device and the results are compared with the traditional RTD device with a realistic ground state transmission line width of $KT/2$, where $T$ is the temperature and $K$ is the Boltzmann constant. This device shows an excellent power-efficiency trade-off when compared to the quantum dot (QD) \cite{BMgriffoni} or the regular RTD based thermoelectric \cite{akshaybm}, and it is possible to achieve high power densities of $0.8 MW/m^{2}$ (twice that of RTD) at an efficiency of $50\%$ of the Carnot efficiency $\eta_{C}$. We also analyze the detailed physics of charge and heat transport through the device and their effects on power output and efficiency. Finally, a comparative study of RTD and ARC-RTD devices is portrayed in terms of power-efficiency trade-off and is concluded on the benefit notes of ARC-RTD devices.\\
\indent A schematic of the device which we have used for simulation is depicted in Fig.~\ref{1}(a). Ideally, it extends to infinity in transverse direction and is of finite length in the growth direction which is also the transport direction. The active channel region consists of an RTD section in between two ARC layers. We use a GaAs well of width $w=4.2nm$ in between two $Al_{x}Ga_{1-x}As$ barriers of width $b=2.4nm$ each for the RTD section, where the aluminum mole-fraction is adjusted to obtain a conduction band offset of $h=0.3eV$. These design parameters ensure a ground state transmission bandwidth of $KT/2$ in the absence of ARC layers. The ARC layer comprises of a AlGaAs barrier of width $b/2$ and height $h$ in between the two GaAs wells of $w/2$ each. The GaAs/AlGaAs material system is chosen due to their excellent match in lattice constants and effective mass, minimal strain and less variability over a wide range of composition \cite{akshaybm}. The device is fairly accurately modeled using a simple nearest neighbor tight binding Hamiltonian with a one band effective mass approximation \cite{akshaybm}.
%\begin{center}
\begin{figure}[!htb]
	\subfigure[]{\includegraphics[height=0.225\textwidth,width=0.225\textwidth]{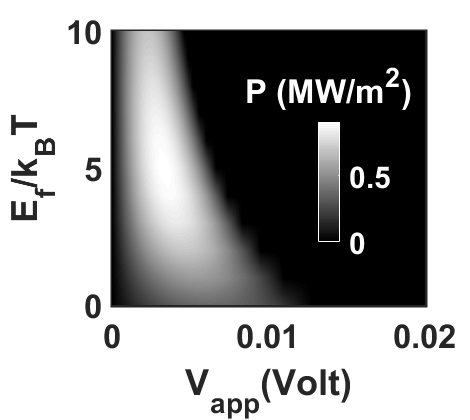}\label{2a}}
	\quad
	\subfigure[]{\includegraphics[height=0.225\textwidth,width=0.225\textwidth]{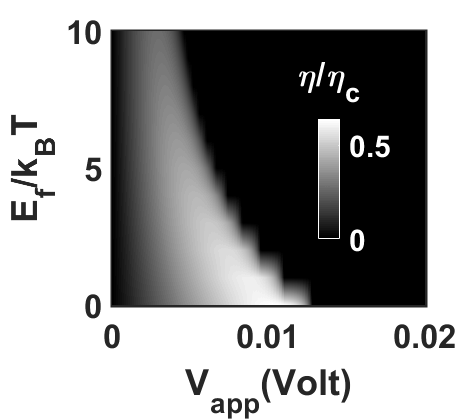}\label{2b}}
	\quad
	
	\caption{Power-efficiency characteristics of the ARC-RTD device: (a) Power density in $MW/m^{2}$ (b) Efficiency $\eta $ (as a fraction of Carnot efficiency $\eta_{c}$) as a function of applied bias for different values of Fermi levels $(E_{f})$ in the units of $KT$. This device generates a maximum power of $0.9 MW/m^{2}$ at $E_{f}=5 KT$ and a maximum efficiency of $65\%$ of $\eta_{c}$ at $E_{f}=0 KT$. }
	\label{2}
\end{figure}
%\end{center}
The schematic of the conduction band diagram along with the transmission spectrum is shown in Fig.~\ref{1}(b). Unlike the RTD or a finite period superlattice, this device produces an almost unity transmission of finite width, which appreciably boosts the transmissivity in the low energy range. The presence of a small dip in transmission is justified due to the slight mismatch in the neighborhood of $E_{m}$.\\
\indent The self-consistent NEGF-Poisson formalism \cite{DattaQT,akshaybm} is employed here within the ballistic regime of transport to calculate the transmission spectrum.  In order to analyze the device performance, we vary the equilibrium Fermi level of the hot and cold contacts, $\mu_{H}$ and $\mu_{C}$ respectively, keeping all other design parameters same. The applied bias ($V_{app}$) shifts the Fermi level of both the hot and cold contacts by an amount of $\pm qV_{app}/2$, where $q$ is the electronic charge unit. The self-consistent solution leads to a non-equilibrium shift of the device transmission function for every change in the contact Fermi level, and the resultant transmission is fed into Landauer equations for charge ($I$) and heat current ($I_{H}^{Q}$) calculation \cite{DattaQT} as follows:   
\begin{equation}
I=\frac{q}{2 \pi \hbar} \sum_{\vec{k}_{\perp}} \int dE T(E) [f_{0}(E+\epsilon_{\vec{k}_{\perp}}-\mu_{H})-f_{0}(E+\epsilon_{\vec{k}_{\perp}}-\mu_{C})], 
\end{equation}
\\and
\begin{multline}
I_{H}^{Q}=\frac{1}{2 \pi \hbar} \sum_{\vec{k}_{\perp}} \int dE T(E) (E+\epsilon_{\vec{k}_{\perp}}-\mu_{H}) \\ \times [f_{0}(E+\epsilon_{\vec{k}_{\perp}}-\mu_{H})-f_{0}(E+\epsilon_{\vec{k}_{\perp}}-\mu_{C})].
\end{multline}
Here, $\hbar$ is the reduced Planck's constant, $K$ is the Boltzmann constant, $T$ is the temperature and $m^{\star}$ is the electron effective mass which is assumed to be uniform throughout the lattice. For our calculation, effective mass is taken as $m^{\star}=0.07m_0$, where $m_0$ is the free electron mass. $T(E)$ is the electronic transmission as a function of energy ($E$) along the transport direction and $\epsilon_{\vec{k}_{\perp}}$ is the assumed parabolic dispersion in the transverse direction represented by wave numbers $\vec{k}_{\perp}$. The contacts are in equilibrium with $f_0(E)$ being the Fermi-Dirac distribution function.  It is important to note that the heat current has two components. The summation over all the transverse momentum ($\vec{k}_{\perp}$) eigenstates is performed with a parabolic dispersion relation assuming periodic boundary conditions. The expressions for the related current densities $J$ simplify to
\begin{equation}
J=\frac{q}{\pi \hbar} \int dE T(E) [f_{2D}(E-\mu_{H})-f_{2D}(E-\mu_{C})], 
\end{equation}
\\and $J_{H}^{Q}=J_{H}^{Q1}+J_{H}^{Q1}$, where $J_{H}^{Q1}$ and $J_{H}^{Q2}$ are given by 
\begin{equation}
J_{H}^{Q1}=\frac{1}{\pi \hbar} \int dE T(E) (E-\mu_{H}) [f_{2D}(E-\mu_{H})-f_{2D}(E-\mu_{C})], 
\end{equation}
\begin{equation}
J_{H}^{Q2}=\frac{1}{\pi \hbar} \int dE T(E) [g_{2D}(E-\mu_{H})-g_{2D}(E-\mu_{C})], 
\end{equation}
\\where,
\begin{equation*}
f_{2D}(E-\mu)=\frac{m^{\star}KT}{2\pi \hbar^{2}} \log(1+e^{(\frac{\mu-E}{KT})}) 
\end{equation*}
\begin{equation*}
g_{2D}(E-\mu)=\frac{m^{\star}KT}{2\pi \hbar^{2}} \int_{0}^{\infty} \frac{\epsilon_{\vec{k}_{\perp}} d\epsilon_{\vec{k}_{\perp}}}{1+\exp(\frac{E+\epsilon_{\vec{k}_{\perp}}-\mu}{KT})}.
\end{equation*}
While the first term ($J_{H}^{Q1}$) is the energy current, the second term ($J_{H}^{Q2}$) depends on the transverse component of the energy. This transverse component is absent in case of a perfect energy filter like a quantum dot but becomes significant for bulk thermoelectric devices \cite{akshaybm}. Having obtained the charge and heat currents, we can evaluate the power output and efficiency using the standard voltage controlled thermoelectric generator set up \cite{BMgriffoni,De2016} with the output power defined as $P=I \times V_{app}$ and the efficiency defined as  $\eta=P/I^Q_{H}$. \\
\indent The power density, that is, the output power per unit cross sectional area, and the efficiency as a function of applied bias for different positions of contact Fermi level are shown in Fig.~\ref{2}(a) and Fig.~\ref{2}(b), respectively. It is seen that the power increases with increasing Fermi level, however, the efficiency decreases for the same. The power attains a maximum of $0.9 MW/m^{2}$ at $E_{f}=5 KT$ and again decreases after that, whereas the maximum efficiency of $65\%$ is attained at $E_{f}=0 KT$. This can be explained via the energy resolved charge and heat currents flowing through the device. The difference of the contact Fermi-Dirac distribution functions crosses the zero mark at an energy ($E_{0}$)
\begin{equation*}
E_{0}=\frac{\mu_{C} T_{H}-\mu_{H} T_{C}-q(V/2)(T_{H}+T_{C})}{(T_{H}-T_{C})}.
\end{equation*}
\begin{center}
	\begin{figure}[!ht]
		\includegraphics[scale=0.3]{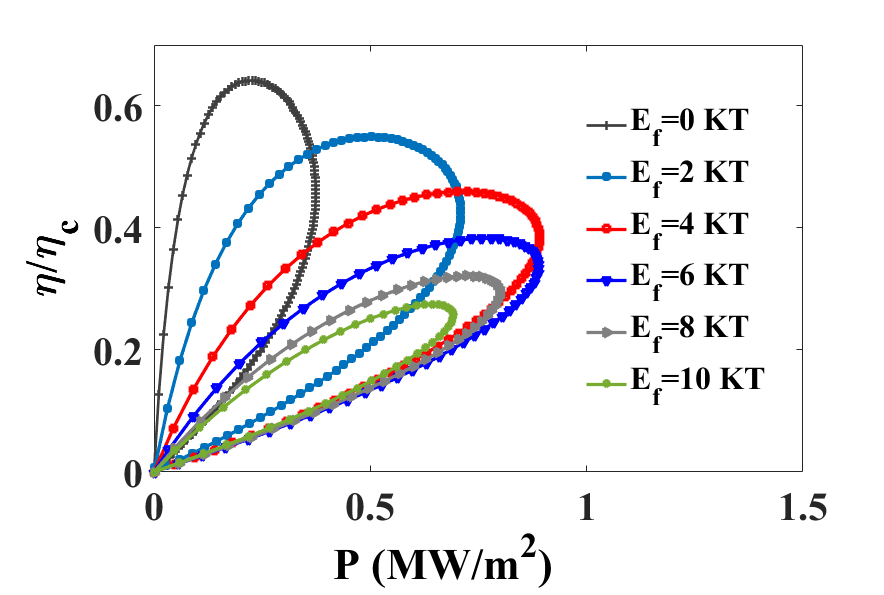}
		\caption{Power-efficiency tradeoffs: A plot of power density versus efficiency $\eta $ (as a fraction of the Carnot efficiency $\eta_{C}$) is shown for different values of Fermi levels. An optimum power efficiency trade-off is achieved in the range of $E_{f}=2$ to $4 KT$ with a power between $0.7$ to $0.9 MW/m^{2}$ and an efficiency between $46$ to $54\%$ of $\eta_{C}$.}
		\label{3}
	\end{figure}
\end{center} 
States above this energy are responsible for electronic flow from hot to cold contact, while the ones below are responsible for the reverse flow. Therefore, it is recommended to have a majority of the states above this energy to ensure a net unidirectional flow of carriers from hot to cold contact. The power becomes maximum when a majority of the states lie above and within a few $KT$ of this energy. However that also tends to increase the heat flow in the same direction and consequently decreases the efficiency. When the Fermi level goes beyond $5KT$, a few of the states move below $E_{0}$ and result in a flow of carriers in the opposite direction, which accounts for a reduction in the output power. At $E_{f}=0KT$, the energy current becomes negligibly small as the DOS lies at a low energy range and the difference in occupation numbers of the contacts becomes negligibly small around the DOS, thereby maximizing the efficiency. \\
\indent The power-efficiency trade-off curves are shown in Fig.~\ref{3} for different positions of Fermi level. An optimum and considerable trade-off is achieved in the range of $E_{f}=2$ to $4 KT$ with a maximum power between $0.7$ to $0.9 MW/m^{2}$ and an efficiency between $46$ to $54\%$ of $\eta_{C}$. As the Fermi level goes beyond $5KT$, both power and efficiency both reduce, which makes this a forbidden operating region.\\
%\begin{center}
\begin{figure}[!ht]
	\subfigure[]{\includegraphics[height=0.225\textwidth,width=0.225\textwidth]{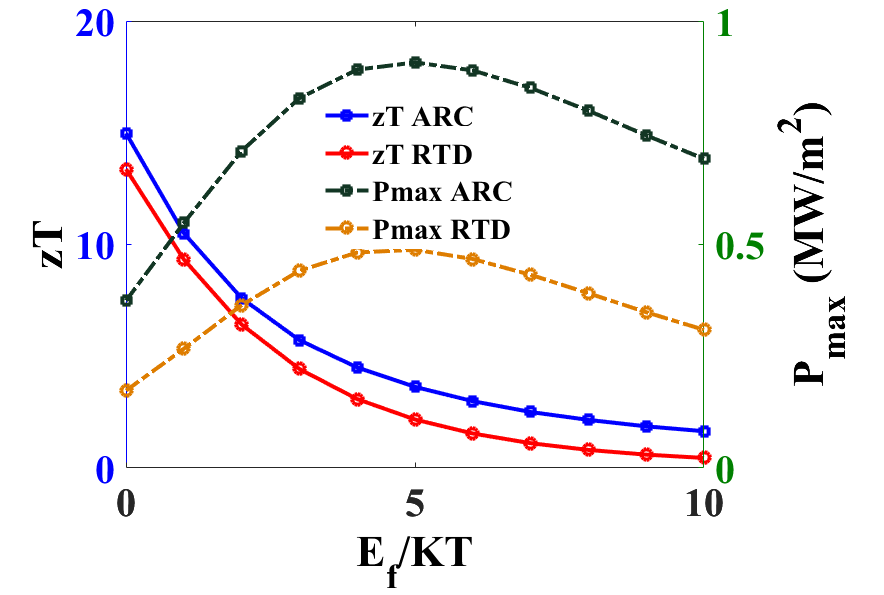}\label{4a}}
	\quad
	\subfigure[]{\includegraphics[height=0.237\textwidth,width=0.225\textwidth]{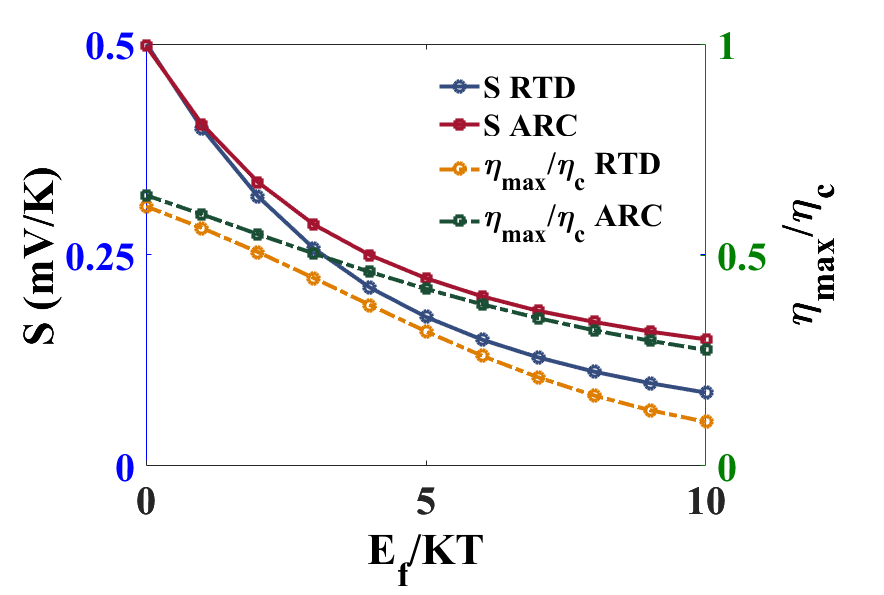}\label{4b}}
	\quad
	\caption{Comparative analysis: (a) $zT$ and maximum power density is shown as a function of different $E_{f}$ for simple RTD thermoelectric and RTD with ARC. A maximum $zT$ of 15 is obtained for the ARC-RTD structure compared to 13 for the regular RTD structure at $E_{f}=0KT$. However the maximum power of the ARC-RTD device becomes twice that of a regular RTD at $E_{f}=5KT$. (b) Seebeck coefficient and maximum efficiency are also shown with varying $E_{f}$. An improvement in both quantities is seen at higher values of $E_{f}$.}
	\label{4}
\end{figure}
%\end{center}
%\begin{center}
\begin{figure}[!ht]
	\subfigure[]{\includegraphics[height=0.235\textwidth,width=0.225\textwidth]{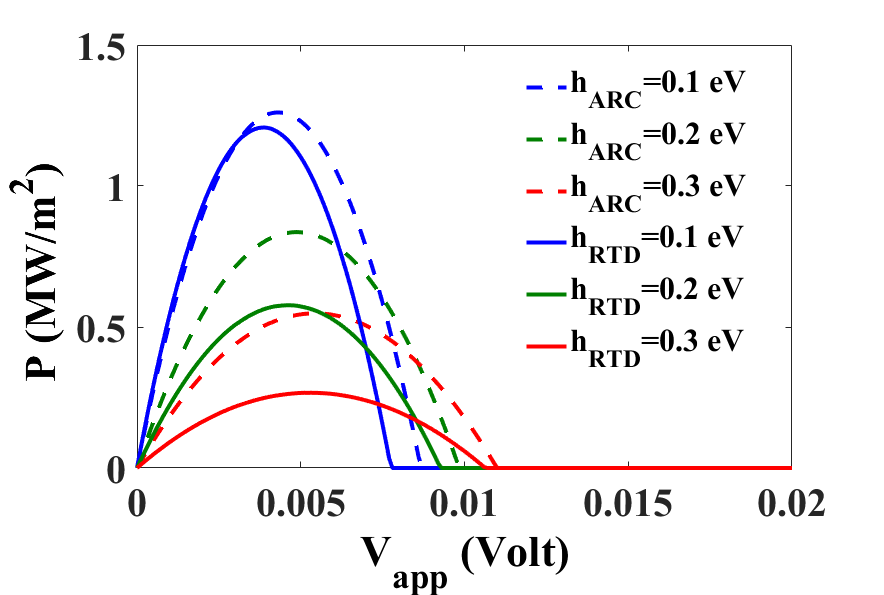}\label{5a}}
	\quad
	\subfigure[]{\includegraphics[height=0.235\textwidth,width=0.225\textwidth]{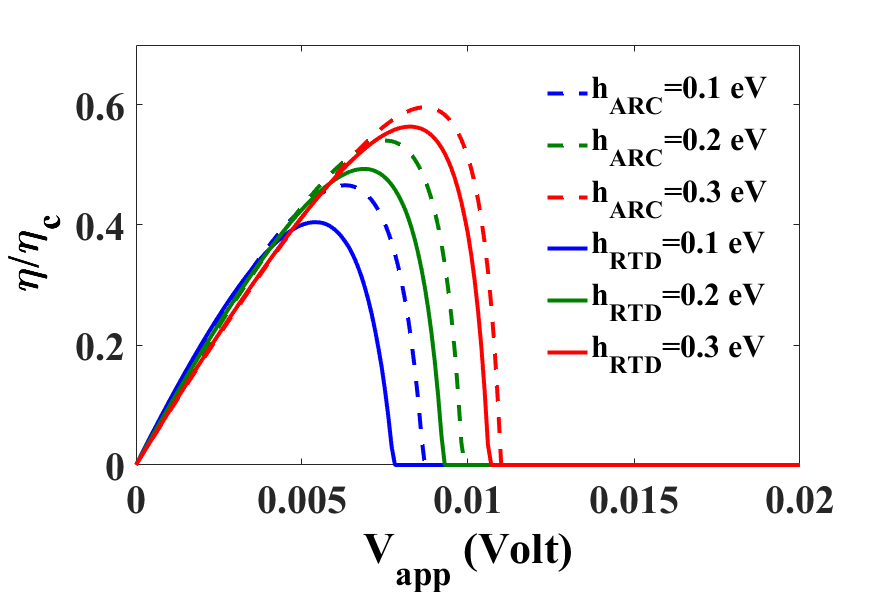}\label{5b}}
	\quad
	\caption{Comparative analysis:  (a) Power and (b) efficiency as a function of applied bias for different values of barrier heights between ARC-RTD (dashed) and the regular RTD (solid) thermoelectric. The ARC-RTD performs far better for taller barriers, but does not improve much for shorter barriers. }
	\label{5}
\end{figure}
%\end{center}
\indent We have also compared the ARC-RTD device with the regular RTD device in terms of all the standard performance parameters. The linear response regime has been considered for this calculation with an assumption of a small temperature gradient between the two contacts. For various Fermi level positions, the figure of merit $zT$ and the maximum power are plotted in Fig.~\ref{4}(a). The plots clearly depict that the maximum power in the ARC-RTD device is favorably enhanced in comparison with the regular RTD device and almost doubles in the mid $E_{f}$ region. The figure of merit, $zT$, is also slightly improved for all values of $E_{f}$. However, the maximum values of $zT$ and power output occurs at different values of $E_{f}$ which clearly points to the power-efficiency trade-off. A comparison of the Seebeck coefficient and the maximum efficiency is also shown in Fig.~\ref{4}(b), which reveals that both these parameters display similar trends at smaller values of Fermi levels, but improve considerably in the case of the ARC-RTD device with increasing Fermi energy.\\
\indent To analyze the results properly, physical phenomena behind them must be clearly understood. The widened transmission spectrum of the ARC-RTD device spans a large number of transverse current carrying modes which boost the output power. On the contrary, the sharp transmission peak of the regular RTD device reduces the number of effective contributing modes. To comment on $zT$ or the efficiency, it is necessary to consider the energy resolved heat currents as well. The non-equilibrium transmission spectrum is pushed up in the energy scale with increasing contact Fermi levels, which effectively increases the heat currents. The parasitic heat current outside the main transmission window becomes less dominant in the case of the ARC-RTD device due to the steep transition of the transmission spectrum, unlike the regular RTD device, which has a broadened nature. At high energies, this effect becomes less detrimental in the ARC-RTD case in terms of efficiencies and makes it more competent in comparison with regular RTD structures.\\
\indent In order to study the effect of broadening on the output power and efficiency, we have varied the barrier heights, which can be realized by varying the mole fraction of Al in the AlGaAs layer. Comparative studies of the ouput power and efficiency are presented in Fig.~\ref{5}(a) and Fig.~\ref{5}(b), respectively. The results clearly suggest that, with decreasing barrier height, the transmission width increases which clearly improves the power, whereas the efficiency deteriorates due to its broadened nature and naturally the reverse effect is observed with increasing barrier height. At shorter barrier heights, although the power of the ARC-RTD device improves remarkably, the margin of improvement remains small in comparison with the regular RTD device. However, the same condition accounts for a smaller value of the efficiency with a higher improvement margin with respect to the regular RTD device. A complete reverse effect is observed for taller barrier heights. It is therefore advisable to set the barrier height to a moderate value to ensure adequate operating power and efficiency to ensure a high margin of improvement over the regular RTD device. \\
\indent In conclusion, we have theoretically analyzed the thermoelectric performance of the ARC-RTD thermoelectric and found that this device significantly prevails over the regular RTD or QD thermoelectric. It is observed that the enabling of ARC almost doubles the output power along with a noticeable improvement in efficiency at its best operating regime. The study however is limited to the ballistic regime of operation, which is a valid assumption in this device dimension, can further be extended to investigate the effect of dephasing processes. Contributions of phonon thermal conductivity can also be considered for a complete understanding of best nano-scale thermoelectric. \\
{\it{Acknowledgements:}} This work was supported in part by the IIT Bombay SEED grant and the Indian Space Research Organization (RESPOND) grant. The authors acknowledge useful discussions with Abhishek Sharma. \\ 
\bibliographystyle{apl}

\bibliography{aip_apl}
\end{document}